\documentclass{kluwer}

\begin{document}      

\begin{article}                                                     
\begin{opening}         
\title{TILTED BIANCHI TYPE V  BULK VISCOUS COSMOLOGICAL 
MODELS  IN GENERAL RELATIVITY}
\author{ANIRUDH PRADHAN$^{1}$\thanks{Corresponding Author}}
\runningtitle{TILTED BIANCHI TYPE V  BULK VISCOUS COSMOLOGICAL MODELS}
\runningauthor{A. PPRADHAN AND ABHA RAI}
\author{ABHA RAI$^{2}$}
\institute{$^{1,2}$ Department of Mathematics, Hindu Post-graduate College,
Zamania-232 331, Ghazipur, U. P., India; \\
E-mail:pradhan@iucaa.ernet.in, acpradhan@yahoo.com}

\institute{$^{2}$ S-25/34-12, Sarsauli Cantt. Varanasi-221 002, India}

\date{\today}

\begin{abstract} 
Conformally flat tilted Bianchi type V cosmological models in presence of 
a bulk viscous fluid and heat flow are investigated. The coefficient of 
bulk viscosity is assumed to be a power function of mass density. Some 
physical and geometric aspects of the models are also discussed.
\end{abstract}

\end{opening}

\section{Introduction}
\vspace*{-0.5pt}
\noindent
A considerable interest has been shown to the study of physical properties of 
spacetimes which are conformal to certain well known gravitational fields. The 
general theory of relativity is believed by a number of unknown functions - the 
ten components of $g^{ij}$. Hence there is a little hope of finding physically 
interesting results without making reduction in their number. In conformally 
flat spacetime the number of unknown functions is reduced to one. The 
conformally flat metrices are of particular interest in view of their 
degeneracy in the context of Petrov classification. A number of conformally  
flat physically significant spacetimes are known like Schwarzschild interior 
solution and Lema\^{i}tre cosmological universe.\\

The study of Bianchi type V cosmological models create more interest as these 
models contain isotropic special cases and permit arbitrarily small 
anisotropy levels at any instant of cosmic time. This property makes them
suitable as model of our universe. Also Bianchi type V models are more 
complicated than the simplest Bianchi type models e.g. the Einstein tensor has 
off diagonal terms so that it is more natural to include tilt and heat conduction.
Spacetimes of Bianchi type I, V and IX universes are the generalizations of
FRW models and it will be interesting to construct cosmological models of
these types which are of class one. Roy and Prasad (1994) have investigated 
Bianchi type V universes which are locally rotationally symmetric and are
of embedding class one filled with perfect fluid with heat conduction and
radiation. Bianchi type V cosmological models have been studied by other
researchers (Farnsworth, 1967; Maarteins and Nel, 1978; Wainwright {\it it al.}
, 1979; Collins, 1974; Meena and Bali, 2002). \\ 

The general dynamics of tilted models have been studied by King and Ellis 
(1973) and Ellis and King (1974). The cosmological models with heat flow 
have been also studied by Coley and Tupper (1983. 1984); Roy and Banerjee 
(1988). Ellis and Baldwin (1984) have shown that we are likely to be living 
in a tilted universe and they have indicated how we may detect it. Beesham
(1986) derived tilted Bianchi type V cosmological models in the scale-covariant
theory. A tilted cold dark matter cosmological scenario has been discussed
by Cen, Nickolay, Kofman and Ostriker (1992). Recently Bali and Meena (2002)
have investigated two tilted cosmological models filled with disordered radiation
of perfect fluid and heat flow. Tilted Bianchi type I cosmological model for 
perfect fluid distribution in presence of magnetic field is investigated by
Bali and Sharma (2003). Tilted Bianchi type I cosmological models filled with
disordered radiation in presence of a bulk viscous fluid and heat flow are 
obtained by Pradhan and Rai (2003).\\

Most cosmological models assume that the matter in the universe can be described 
by `dust'(a pressureless distribution) or at best a perfect fluid. Nevertheless,
there is good reason to believe that - at least at the early stages of the universe 
- viscous effects do play a role (Israel and Vardalas, 1970; Klimek, 1971;
Weinberge, 1971). For example, the existence of the bulk viscosity is equivalent to 
slow process of restoring equilibrium states (Landau and Lifshitz, 1962). The 
observed physical phenomena such as the large entropy per baryon and remarkable 
degree of isotropy of the cosmic microwave background radiation suggest analysis 
of dissipative effects in cosmology. Bulk viscosity is associated with 
the GUT phase transition and string creation. Thus, we should consider the 
presence of a material distribution other than a perfect fluid to have realistic 
cosmological models (see Gr\o n ) for a review on cosmological models with bulk 
viscosity). The model studied by Murphy (1973) possessed an interesting feature 
in that the big bang type of singularity of infinite spacetime curvature does not 
occur to be a finite past. However, the relationship assumed by Murphy between the
viscosity coefficient and the matter density is not acceptable at large 
density. The effect of bulk viscosity on the cosmological evolution has been 
investigated by a number of authors in the framework of general theory of 
relativity .(Pavon, 1991; Padmanabhan and Chitre, 1987; Johri and Sudarshan, 1988; 
Maartens, 1995; Zimdahl, 1996; Santos {\it et al.}, 1985; Pradhan, Sarayakar and 
Beesham, 1997; Kalyani and Singh 1997; Singh, Beesham and Mbokazi, 1998; 
Pradhan {\it et al.}, 2001, 2002, 2003). This motivates to study cosmological bulk 
viscous fluid model.\\  
Recently Meena and Bali (2002) have investigated two conformally flat tilted 
Bianchi type V cosmological models filled with a perfect fluid and heat conduction. 
In this paper, we propose to find tilted Bianchi type V cosmological 
models in presence of a bulk viscous fluid and heat flow.\\ 
\section{The metric and field  equations}
We consider the Bianchi type V metric in the form
\begin{equation}
\label{eq1}
ds^{2} = - dt^{2} + A^{2} dx^{2} + B^{2} e^{2x} \left(dy^{2} + dz^{2}\right),
\end{equation}
where A, B are function of $t$ only.\\
The Einstein's field equations (in gravitational units $c = 1$, $G = 1$) read as
\begin{equation}
\label{eq2}
R^{j}_{i} - \frac{1}{2} R g^{j}_{i} = -8\pi T^{j}_{i},
\end{equation}
where $R^{j}_{i}$ is the Ricci tensor; $R$ = $g^{ij} R_{ij}$ is the
Ricci scalar; and $T^{j}_{i}$ is the stress energy-tensor in the presence
of bulk stress given by 
\begin{equation}
\label{eq3}
T^{j}_{i} = (\rho + \bar{p})v_{i}v^{j} + \bar{p} g^{j}_{i} + 
q_{i}v^{j} + v_{i}q^{j},
\end{equation}
and
\begin{equation}
\label{eq4}
\bar{p} = p - \xi v^{i}_{;i}.
\end{equation}
Here $\rho$, $p$, $\bar{p}$ and $\xi$ are the energy density,
isotropic pressure, effective pressure and  bulk viscous 
coefficient respectively and $v_{i}$ is the flow vector satisfying 
the relations
\begin{equation}
\label{eq5}
g_{ij} v^{i}v^{j} = - 1,
\end{equation}
\begin{equation}
\label{eq6}
q_{i} q^{j} > 0,
\end{equation}
\begin{equation}
\label{eq7}
q_{i}v^{i} = 0,
\end{equation}
where $q_{i}$ is the  heat conduction vector orthogonal to $v_{i}$.
The fluid flow vector has the components $(\frac{\sinh \lambda}{A}, 0, 0, 
\cosh \lambda)$ satisfying Eq. (\ref{eq5}) and $\lambda$ is the tilt angle.\\
The Einstein's field equations (\ref{eq2}) for the line element (\ref{eq1})
has been set up as
\begin{equation}
\label{eq8}
- 8\pi[(\rho + \bar{p})\sinh^{2}  \lambda + \bar{p} + 2 A q_{1} \sinh \lambda] 
= \frac{2 B_{44}}{B} + \left(\frac{B_{4}}{B}\right)^{2} - \frac{1}{A^{2}}, 
\end{equation}
\begin{equation}
\label{eq9}
- 8\pi \bar{p} = \frac{A_{44}}{A} + \frac{B_{44}}{B} + \frac{A_{4}B_{4}}{AB}
- \frac{1}{A^{2}}, 
\end{equation}
\begin{equation}
\label{eq10}
- 8\pi[- (\rho + \bar{p})\cosh^{2} \lambda + \bar{p} - 2 A q_{1} \sinh \lambda
] = \frac{2 A_{4}B_{4}}{AB} + \left(\frac{B_{4}}{B}\right)^{2} - \frac{3}{A^{2}},  
\end{equation}
\begin{equation}
\label{eq11}
-8 \pi[(\rho + \bar{p})A \sinh \lambda ~  \cosh \lambda + A^{2} q_{1}
(\cosh \lambda + \sinh \lambda ~ \tanh \lambda)] = \frac{2 A_{4}}{A}
- \frac{2 B_{4}}{B},
\end{equation}
where the suffix $4$ at the symbols $A$, $B$ denotes ordinary 
differentiation with respect to $t$.
\section{Solution of the field equations}
Equations (\ref{eq8}) - (\ref{eq12}) are four independent equations in seven 
unknowns $A$, $B$, $\rho$, $p$, $\xi$, $q$ and $\lambda$. For the complete 
determinacy of the system, we need three extra conditions.\\
First we assume that the spacetime is conformally flat which leads to
\begin{equation}
\label{eq12}
C_{2323} = \frac{1}{3}\left[\frac{A_{44}}{A} - \frac{B_{44}}{B} - \frac{A_{4}B_{4}}
{AB} + \frac{B^{2}_{4}}{B^{2}}\right] = 0
\end{equation}
and secondly, we assume 
\begin{equation}
\label{eq13}
A = B^{n},
\end{equation}
where $n$ is any  real number.
Eqs. (\ref{eq12}) and (\ref{eq13}) lead to
\begin{equation}
\label{eq14}
\frac{B_{44}}{B} + (n - 1)\frac{B^{2}_{4}}{B^{2}} = 0.
\end{equation}
From  Equations (\ref{eq8}), (\ref{eq10}) and (\ref{eq13}), we have 
\begin{equation}
\label{eq15}
- 4 \pi \left[(\rho + \bar{p}) \cosh 2\lambda + 4 B^{n} q_{1} \sinh \lambda\right]
= \frac{B_{44}}{B} - n \frac{B^{2}_{4}}{B^{2}} + \frac{1}{B^{2n}},
\end{equation}
and
\begin{equation}
\label{eq16}
4 \pi (\rho - \bar{p}) = \frac{B_{44}}{B} + (n + 1) \frac{B^{2}_{4}}{B^{2}} 
- \frac{2}{B^{2n}}.
\end{equation}
Equations (\ref{eq9}), (\ref{eq16})  and (\ref{eq12}) lead to
\begin{equation}
\label{eq17}
 (n - n^{2} + 1)\frac{B^{2}_{4}}{B^{2}} - n \frac{B_{44}}{B} - 
\frac{1}{B^{2n}} = 4 \pi (\rho + \bar{p}).
\end{equation}
Equations (\ref{eq11}) and (\ref{eq13}) lead to
\begin{equation}
\label{eq18}
- 16 \pi q_{1} B^{n} \sinh \lambda = \frac{2(n - 1)B_{4} \tanh 2\lambda}
{B^{n + 1}} + 4\pi (\rho + \bar{p})\sinh 2\lambda~ \tanh 2\lambda.
\end{equation}
From Eqs. (\ref{eq15}) and (\ref{eq18}), we obtain 
\begin{equation}
\label{eq19}
\frac{B_{44}}{B} - \frac{n B^{2}_{4}}{B} +
\frac{1}{B^{2n}} = - \frac{4\pi(\rho + \bar{p})}{\cosh 2\lambda} +
\frac{2(n - 1) \tanh 2\lambda}{B^{n}}.
\end{equation}
Equations (\ref{eq17}) and (\ref{eq19}) lead to 
\begin{equation}
\label{eq20}
\frac{B_{44}}{B} - \frac{n B^{2}_{4}}{B} + \frac{1}{B^{2n}} =
\frac{n B_{44}}{B} + \frac{(n^{2} - n - 1) B^{2}_{4}}{B^{2}} + B^{2n}{\rm sech}
2\lambda + \frac{2(n - 1)~ B_{4}\tanh 2\lambda}{B^{n + 1}}.
\end{equation}
Equation (\ref{eq14}) can be rewritten as
\begin{equation}
\label{eq21}
\frac{B_{44}}{B_{4}} + \frac{(n - 1) B_{4}}{B} = 0,
\end{equation}
which on integration leads to
\begin{equation}
\label{eq22}
B = n^{\frac{1}{n}} ~ (\alpha t + \beta)^{\frac{1}{n}}.
\end{equation}
where $\alpha$, $\beta$ are constants of integration. Hence we obtain
\begin{equation}
\label{eq23}
A^{2} = n^{2} ~ (\alpha t + \beta)^{2},
\end{equation}
\begin{equation}
\label{eq24}
B^{2} = n^{\frac{2}{n}} ~ (\alpha t + \beta)^{\frac{2}{n}}.
\end{equation}
Hence the geometry of the spacetime (\ref{eq1}) reduces to the form
\begin{equation}
\label{eq25}
ds^{2} = - dt^{2} + n^{2} (\alpha t + \beta)^{2} dx^{2} + [n(\alpha t + \beta)]^
{\frac{2}{n}} e^{2x} (dy^{2} + dz^{2}).
\end{equation}
After the suitable transformation of coordinates, the metric (\ref{eq25}) 
takes the form
\begin{equation}
\label{eq26}
ds^{2} = - \frac{dT^{2}}{2} + n^{2} T^{2} dX^{2} + n^{\frac{2}{n}}
T^{\frac{2}{n}} e^{2X}(dY^{2} + dZ^{2}).
\end{equation}
The effective pressure and density of the model (\ref{eq26}) are given by
\begin{equation}
\label{eq27}
8\pi \bar{p} = 8\pi (p -\xi \theta) = - \frac{(\alpha^{2} - 1)}{n^{2} T^{2}},
\end{equation} 
\begin{equation}
\label{eq28}
8\pi \rho = \frac{3(\alpha^{2} - 1)}{n^{2} T^{2}},
\end{equation} 
where $\theta$ is the scalar of expansion calculated for the flow vector 
$v^{i}$ and given is as
\begin{equation}
\label{eq29}
\theta = \frac{2K + (n + 2)k\alpha}{nT}.
\end{equation} 
The tilt angle $\lambda$ is given by
\begin{equation}
\label{eq30}
\cosh^{2}  \lambda = k^{2},
\end{equation} 
\begin{equation}
\label{eq31}
\sinh^{2}  \lambda = K^{2},
\end{equation} 
where $k$ and $K$ are constants given by
\begin{equation}
\label{eq32}
k^{2} = \frac{(n\alpha^{2} - 1)^{2}}{(\alpha^{2} - 1) (2n^{2} \alpha^{2} 
- n \alpha^{2} - 1) + 2\alpha^{2} (n - 1)\sqrt{(n^{2} - n)(1 - \alpha^{2})}},
\end{equation} 
\begin{equation}
\label{eq33}
K^{2} = \frac{(n - 1)(2n \alpha^{2} - n\alpha^{4} - \alpha^{2})
- 2\alpha^{2} (n - 1)\sqrt{(n^{2} - n)(1 - \alpha^{2})}}{(\alpha^{2} - 1) 
(2n^{2} \alpha^{2} - n \alpha^{2} - 1) + 2\alpha^{2} (n - 1)\sqrt{(n^{2} - n)
(1 - \alpha^{2})}}.
\end{equation} 
If we put $\xi = 0$ in (\ref{eq27}), we get the solutions
as obtained by Meena and Bali (2002).\\
Thus, given $\xi(t)$ we can solve the system for the physical quantities.
Therefore to apply the third condition, let us assume the following {\it adhoc}
law (Maartens, 1995; Zimdahl, 1996) 
\begin{equation}
\label{eq34}
\xi(t) = \xi_{0} \rho^{m}
\end{equation} 
where $\xi_{0}$ and $m$ are real constants. If $m = 1$, Eq. (\ref{eq34})
may correspond to a radiative fluid (Weinberg, 1972), whereas 
$m$ = $\frac{3}{2}$ may correspond to a string-dominated universe. However, 
more realistic models (Santos, 1985) are based on lying the regime 
$0 \leq m \leq \frac{1}{2}$. 
\subsection {Model I: ~ ~ ~ $(\xi = \xi_{0})$}
When $m = 0$, Equation (\ref{eq34}) reduces to $\xi = \xi_{0}$ = constant and hence 
Equation (\ref{eq27}) with the use of (\ref{eq29}) leads to
\begin{equation}
\label{eq35}
p = \frac{\xi_{0} \{2K + (n + 2) k \alpha \}}{n T} - \frac{(\alpha^{2} - 1)}
{8 \pi n^{2} T^{2}}.
\end{equation} 
\subsection {Model II: ~ ~ ~ $(\xi = \xi_{0}\rho)$}
When $m = 1$, Equation (\ref{eq34}) reduces to $\xi = \xi_{0}\rho$ and hence 
Equation (\ref{eq27}) with the use of (\ref{eq29}) leads to
\begin{equation}
\label{eq36}
p = \frac{1}{8 \pi n^{2} T^{2}}\left[1 - \alpha^{2}  + \frac{3\xi_{0} 
(\alpha^{2} - 1) \{2K + (n + 2) k \alpha \} }{n T} \right].
\end{equation} 
It is observed from Equations (\ref{eq28}), (\ref{eq35}) and (\ref{eq36})
that $\rho$ and $p$ vary as $\frac{1}{T}$. The models are singular at 
$T = 0$ and as they evolve, the pressure and density decrease.
\section{Some Physical and Geometric Properties of the Models}
The weak and strong energy conditions, we have, in Model I
\begin{equation}
\label{eq37}
\rho + p = \frac{(\alpha^{2} - 1)}{4\pi n^{2} T^{2}} + \frac{\xi_{0} \{2K + 
(n + 2) k \alpha \}}{n T},
\end{equation} 
\begin{equation}
\label{eq38}
\rho - p = \frac{(\alpha^{2} - 1)}{2\pi n^{2} T^{2}} - \frac{\xi_{0} \{2K + 
(n + 2) k \alpha \}}{n T},
\end{equation} 
\begin{equation}
\label{eq39}
\rho + 3p = \frac{3 \xi_{0} \{2K + (n + 2) k \alpha \}}{n T},
\end{equation} 
\begin{equation}
\label{eq40}
\rho - 3p = \frac{(3\alpha^{2} - 1)}{4\pi n^{2} T^{2}} - \frac{3\xi_{0} \{2K + 
(n + 2) k \alpha \}}{n T},
\end{equation} 
In Model II, we have
\begin{equation}
\label{eq41}
\rho + p = \frac{1}{8 \pi n^{2} T^{2}}\left[2(\alpha^{2} - 1) + \frac{3\xi_{0} 
(\alpha^{2} - 1)\{2K + (n + 2) k \alpha \}}{n T}\right],
\end{equation} 
\begin{equation}
\label{eq42}
\rho - p = \frac{1}{8 \pi n^{2} T^{2}}\left[4(\alpha^{2} - 1) - \frac{3\xi_{0} 
(\alpha^{2} - 1)\{2K + (n + 2) k \alpha \}}{n T}\right],
\end{equation} 
\begin{equation}
\label{eq43}
\rho + 3p = \frac{9\xi_{0}(\alpha^{2} - 1)\{2K + (n + 2) k \alpha \}}
{8\pi n^{3} T^{3}},
\end{equation} 
\begin{equation}
\label{eq44}
\rho - 3p = \frac{3}{8 \pi n^{2} T^{2}}\left[2(\alpha^{2} - 1) + \frac{3\xi_{0} 
(\alpha^{2} - 1)\{2K + (n + 2) k \alpha \}}{n T}\right].
\end{equation} 
The reality conditions $\rho \geq 0$, $p \geq 0$ and $\rho - 3p  \geq 0$
impose further restrictions on both of these models.\\
The flow vector $v^{i}$ and heat conduction vector $q^{i}$ for the models 
(\ref{eq26}) are obtained by Meena and Bali (2002)
\begin{equation}
\label{eq45}
v^{1} = \frac{K}{n T},
\end{equation} 
\begin{equation}
\label{eq46}
v^{4} = k,
\end{equation} 
\begin{equation}
\label{eq47}
q^{1} = - \frac{k \{(\alpha^{2} - 1)k  K + (n - 1)\alpha\}}
{4\pi n^{3} T^{3} (k^{2} + K^{2})},
\end{equation} 
\begin{equation}
\label{eq48}
q^{4} =  - \frac{K \{(\alpha^{2} - 1)k K + (n - 1)\alpha\}}
{4\pi n^{3} T^{3} (k^{2} + K^{2})}.
\end{equation}  
The rate of expansion $H_{i}$ in the direction of $X$, $Y$, $Z$-axe
are given by
\begin{equation}
\label{eq49}
H_{1} = \frac{\alpha}{T}, 
\end{equation} 
\begin{equation}
\label{eq50}
H_{2} = H_{3} = \frac{\alpha}{n T}.
\end{equation} 
The non-vanishing components of shear tensor $(\sigma_{ij})$ and rotation
tensor $(\omega_{ij})$ are obtained as
\begin{equation}
\label{eq51}
\sigma_{11} = \frac{2}{3} n k^{2} T [(n - 1)k \alpha - K],
\end{equation} 
\begin{equation}
\label{eq52}
\sigma_{22} = \sigma_{33} = \frac{(nT)^{\frac{2}{n} - 1} e^{2X}}{3} 
[(1 - n)k \alpha - 2K],
\end{equation} 
\begin{equation}
\label{eq53}
\sigma_{44} =  \frac{2 K^{2}}{3 n T}[(n - 1)k \alpha - K],
\end{equation} 
\begin{equation}
\label{eq54}
\sigma_{14} = \frac{K}{3}[2(1 - n) k^{2} + 2 k K - 3 n],
\end{equation} 
\begin{equation}
\label{eq55}
\omega_{14} = n \alpha K.
\end{equation} 
The models, in general, represent shearing and  rotating universes. The models
start expanding with a Big bang  at $T = 0$ and the expansion 
in the models decreases as time increases and the expansion in the models stops 
at $T = \infty$ and $\alpha = - \frac{2K}{(n + 2) k}$. Both density and pressure 
in the models become zero at $ T = \infty$. For $\alpha = 1$, $n = 1$, we observe
that heat conduction vector $q^{1} = q^{4} = 0$. When $T \rightarrow \infty$, $v^{1}
= 0$, $v^{4} = $ constant, $q^{1} = q^{4} = 0$. Since $\lim_{t\rightarrow \infty}
\frac{\sigma}{\theta} \ne 0$, the models do not approach isotropy for large values 
of $T$. There is a real physical singularity in the model at $T = 0$.\\
\section{Particular Models }
If we set $n = 2$, then the geometry of the spacetime (\ref{eq26}) reduces to 
the form
\begin{equation}
\label{eq56}
ds^{2} = - \frac{dT^{2}}{2} + T dX^{2} + T e^{X} ( dY^{2} + dZ^{2}).
\end{equation} 
The effective pressure and density for the model (\ref{eq56}) are given by
\begin{equation}
\label{eq57}
8\pi \bar{p} = 8\pi (p -\xi \theta) = - \frac{(\alpha^{2} - 1)}{4 T^{2}},
\end{equation} 
\begin{equation}
\label{eq58}
8\pi \rho = \frac{3(\alpha^{2} - 1)}{4 T^{2}},
\end{equation} 
where $\theta$ is the scalar expansion obtained as
\begin{equation}
\label{eq59}
\theta = \frac{K_{1} + 2k_{1} \alpha}{T}.
\end{equation} 
The tilt angle $\lambda$ is given by
\begin{equation}
\label{eq60}
\cosh^{2}  \lambda = k_{1}^{2},
\end{equation} 
\begin{equation}
\label{eq61}
\sinh^{2}  \lambda = K^{2}_{1},
\end{equation} 
where $k_{1}$ and $K_{1}$ are constants given by
\begin{equation}
\label{eq62}
k^{2}_{1} = \frac{(2 \alpha^{2} - 1)^{2}}{6\alpha^{4} - 7 \alpha^{2} 
+ 1 + 2\alpha^{2} \sqrt{2(1 - \alpha^{2})}},
\end{equation} 
\begin{equation}
\label{eq63}
K^{2}_{1} = \frac{\alpha^{2}\left[(3 - 2\alpha^{2}
-2 \sqrt{2(1 - \alpha^{2})} \right] } {6 \alpha^{4} - 7 \alpha^{2} + 1 + 
2\alpha^{2} \sqrt{2(1 - \alpha^{2})}}.
\end{equation} 
Thus, for given $\xi(t)$ one can solve the system for the physical quantities.
\subsection {Model I: ~ ~ ~ $(\xi = \xi_{0})$}
When $m = 0$, Equation (\ref{eq34}) reduces to $\xi = \xi_{0}$ = constant and hence 
Equation (\ref{eq57}) with the use of (\ref{eq59}) leads to
\begin{equation}
\label{eq64}
p = \frac{\xi_{0} \{K_{1} + 2 k_{1} \alpha \}}{T} - \frac{(\alpha^{2} - 1)}
{32 \pi T^{2}}.
\end{equation} 
\subsection {Model II: ~ ~ ~ $(\xi = \xi_{0}\rho)$}
When $m = 1$, Equation (\ref{eq34}) reduces to $\xi = \xi_{0}\rho$ and hence 
Equation (\ref{eq57}) with the use of (\ref{eq59}) leads to
\begin{equation}
\label{eq65}
p = \frac{1}{32 \pi T^{2}}\left[1 - \alpha^{2}  + \frac{3\xi_{0} 
(\alpha^{2} - 1) \{K_{1} + 2 k_{1} \alpha \} }{ T} \right].
\end{equation} 
It is observed from Equations (\ref{eq58}), (\ref{eq64}) and (\ref{eq65})
that $\rho$ and $p$ vary as $\frac{1}{T}$. The models are singular at 
$T = 0$ and as they evolve, the pressure and density decrease.
\section{Some Physical and Geometric Properties of Particular Models}
The weak and strong energy conditions, we have, in Model I
\begin{equation}
\label{eq66}
\rho + p = \frac{(\alpha^{2} - 1)}{16\pi T^{2}} + \frac{\xi_{0} \{K_{1} + 
2 k_{1} \alpha \}}{ T},
\end{equation} 
\begin{equation}
\label{eq67}
\rho - p = \frac{(\alpha^{2} - 1)}{8\pi T^{2}} - \frac{\xi_{0} \{K_{1} + 
2k_{1} \alpha \}}{T},
\end{equation} 
\begin{equation}
\label{eq68}
\rho + 3p = \frac{3 \xi_{0} \{K_{1} + 2 k_{1} \alpha \}}{ T},
\end{equation} 
\begin{equation}
\label{eq69}
\rho - 3p = \frac{(3\alpha^{2} - 1)}{16 \pi T^{2}} - \frac{3\xi_{0} \{K_{1} + 
2k_{1} \alpha \}}{T},
\end{equation} 
In Model II, we have
\begin{equation}
\label{eq70}
\rho + p = \frac{1}{32 \pi T^{2}}\left[2(\alpha^{2} - 1) + \frac{3\xi_{0} 
(\alpha^{2} - 1)\{K_{1} + 2k_{1} \alpha \}}{T}\right],
\end{equation} 
\begin{equation}
\label{eq71}
\rho - p = \frac{1}{32 \pi T^{2}}\left[4(\alpha^{2} - 1) - \frac{3\xi_{0} 
(\alpha^{2} - 1)\{K_{1} + 2k_{1} \alpha \}}{T}\right],
\end{equation} 
\begin{equation}
\label{eq72}
\rho + 3p = \frac{9\xi_{0}(\alpha^{2} - 1)\{K_{1} + 2k_{1} \alpha \}}
{32\pi n^{3} T^{3}},
\end{equation} 
\begin{equation}
\label{eq73}
\rho - 3p = \frac{3}{32 \pi n^{2} T^{2}}\left[2(\alpha^{2} - 1) + \frac{3\xi_{0} 
(\alpha^{2} - 1)\{K_{1} + 2k_{1} \alpha \}}{T}\right].
\end{equation} 
The reality conditions $\rho \geq 0$, $p \geq 0$ and $\rho - 3p  \geq 0$
impose further restrictions on both of these models.\\
The flow vector $v^{i}$ and heat conduction vector $q^{i}$ for the models 
(\ref{eq56}) are obtained as
\begin{equation}
\label{eq74}
v^{1} = \frac{K_{1}}{2 T},
\end{equation} 
\begin{equation}
\label{eq75}
v^{4} = k_{1},
\end{equation} 
\begin{equation}
\label{eq76}
q^{1} = - \frac{k_{1} \{(\alpha^{2} - 1)k_{1}  K_{1} + \alpha\}}
{32\pi T^{3} (k^{2}_{1} + K^{2}_{1})},
\end{equation} 
\begin{equation}
\label{eq77}
q^{4} =  - \frac{K_{1} \{(\alpha^{2} - 1)k_{1} K_{1} + \alpha\}}
{32\pi T^{3} (k^{2}_{1} + K^{2}_{1})}.
\end{equation}  
The rate of expansion $H_{i}$ in the direction of $X$, $Y$, $Z$-axe
are given by
\begin{equation}
\label{eq78}
H_{1} = \frac{\alpha}{T}, 
\end{equation} 
\begin{equation}
\label{eq79}
H_{2} = H_{3} = \frac{\alpha}{2 T}.
\end{equation} 
The non-vanishing components of shear tensor $(\sigma_{ij})$ and rotation
tensor $(\omega_{ij})$ are obtained as
\begin{equation}
\label{eq80}
\sigma_{11} = \frac{4}{3} k^{2}_{1} T [k_{1} \alpha - K_{1}],
\end{equation} 
\begin{equation}
\label{eq81}
\sigma_{22} = \sigma_{33} = - \frac{e^{2X}}{3} [k_{1} \alpha + 2 K_{1}],
\end{equation} 
\begin{equation}
\label{eq82}
\sigma_{44} =  \frac{K^{2}_{1}}{3 T}[k_{1} \alpha - K_{1}],
\end{equation} 
\begin{equation}
\label{eq83}
\sigma_{14} = - \frac{K_{1}}{3}[2 k^{2}_{1} - 2 k_{1} K_{1} + 6],
\end{equation} 
\begin{equation}
\label{eq84}
\omega_{14} = 2 \alpha K_{1}.
\end{equation} 
The models in general represent shearing and  rotating universes. The models
start expanding with a Big bang  at $T = 0$ and the expansion 
in the models decreases as time increases and the expansion in the models stops 
at $T = \infty$ and $\alpha = - \frac{K_{1}}{2 k_{1}}$. Both density and pressure 
in the models become zero at $ T = \infty$. For expansion and rotation, we have
$\alpha \neq 1$, $K_{1} \neq 1$ when $ T \rightarrow \infty$ we observe
that heat conduction vector $q^{1} = q^{4} = 0$, $v^{1} = 0$, $v^{4} = K_{1}$. 
Since $\lim_{t\rightarrow \infty} \frac{\sigma}{\theta} \ne 0$, the models do 
not approach isotropy for large values of $T$. There is a real physical 
singularity in the model at $T = 0$. When $k_{1} = 1$ then $\lambda = 0$.
Thus, the tilted cosmological models lead to non-tilted one for $k_{1} = 1$.\\ 
\section{Conclusions}
We have obtained a new class of conformally flat tilted Bianchi type V
magnetized cosmological models with a bulk viscous fluid as the source
of matter. Generally, the models are expanding, shearing and rotating. 
In all these models, we observe that they do not approach isotropy for
large values of time $T$ in the presence of magnetic field. \\
The coefficient of bulk viscosity is assumed to be a power function of
mass density. The effect of bulk viscosity is to introduce a change in the 
perfect fluid model. We also observe here that the conclusion of Murphy (1973)
about the absence of a Big bang type of singularity in the finite past in models 
with bulk viscous fluid is, in general, not true.  
\section*{Acknowledgements} 
A. Pradhan thanks to the Inter-University Centre for Astronomy and Astrophysics, 
India for providing  facility under Associateship Programme where the part of 
work was carried out. \\
\newline
\newline

\end{article}
\end{document}